\documentclass[aps,prd,,showpacs,superscriptaddress,groupedaddress]{revtex4}  % for review and submission
\usepackage{graphicx}  % needed for figures
\usepackage{dcolumn}   % needed for some tables
\usepackage{bm}        % for math
\usepackage{amssymb}   % for math
\usepackage{color}

\begin{document}

\title{Elastic vector and axial scattering of weakly interacting particles off nuclei}

\author{O. Moreno}
\author{T. W. Donnelly}
\affiliation{Center for Theoretical Physics, Laboratory for Nuclear Science and Department of Physics, Massachusetts Institute of Technology, Cambridge, MA 02139, USA}
\date{\today}

\begin{abstract}
We analyze the elastic scattering of particles interacting with nuclei through vector and axial currents with overall couplings of the order of the Standard Model weak interaction, or smaller; the dominant contribution to the elastic cross section is identified as the coherent component and is therefore spin-independent. Differential and integrated cross sections are obtained for a wide range of incident particle masses and velocities and for nuclear targets with different masses; vector, axial and overall couplings of the incident particle and of the hadronic target to the massive exchanged boson are also kept general. This study naturally encompasses several kinds of possible dark matter components, including active and sterile neutrinos or neutralinos, and addresses the prospects for their direct detection through elastic scattering off nuclei.
\end{abstract}

\maketitle

\section{Introduction}\label{introduction}

Weakly interacting particles (WIPs) are natural candidates to contribute to the dark matter content of the Universe, which accounts for around 25$\%$ of the total energy density according to growing astrophysical and cosmological evidence \cite{ber05, roos10}. These observations are a consequence of the gravitational interaction of the WIPs, but they are assumed also to participate in other interactions with strengths of the same order of, or smaller than, the Standard Model (SM) weak neutral current (WNC); this fact opens the possibility of their direct detection, but also accounts for its difficulty. Current observations favor  weakly interacting {\it massive} particles (WIMPs), with masses from a few keV (warm dark matter) to several GeV or TeV (cold dark matter), as the {\it main} components of dark matter in the Universe. However, any WIP, independently of its mass, can make a contribution to the dark content whose significance will depend on the specific mass, energy and relic density of such particles.

We study WIPs within a wide range of velocities, or equivalently energies or momenta, as well as of masses. We consider elementary spin-1/2 WIPs interacting through a weak-like current with arbitrary overall strength containing a vector (polar vector) component and an axial (axial vector) component with arbitrary relative weight. We will focus on the direct detection of the so-defined WIPs through their elastic scattering off nuclei. In this process the nature of the incoming and the outgoing WIP is the same and, given its small interaction probability, the scattering can only manifest itself through the recoil of the nuclear target. Direct detection can be complemented with indirect detection, based on the observation of standard particles produced in the annihilation of relic Majorana WIPs. The existence of WIPs can also be made apparent through their production in decays or collisions that lead to missing energy and momentum, to be reconstructed from the measurements under controlled experiments; for the production of very heavy WIPs, experiments at high-energy facilities such as the LHC are the only option.

On the nuclear vertex, we consider an interaction involving the vector nuclear response, which is predominantly spin-independent and where all the nucleons in the target contribute coherently to the elastic scattering cross section. This is in contrast to the spin-dependent scattering, which depends on the nuclear spin to which only one, or at most a few of the nucleons contribute. This is the case when the axial nuclear response is involved, which is more sensitive to the nuclear structure and where both elastic and inelastic proceses are of interest (see, for instance, \cite{men12} for neutralino scattering using chiral effective field theory). Another type of spin-independent interaction is the coupling of WIP-scalar and nuclear-scalar currents, which can be also applied to elastic and inelastic processes \cite{vie15}.

The vector nuclear current, having a coherent component, dominates over the axial one in elastic scattering, and is the only contribution for even-$Z$ even-$N$ nuclei; it will be the only contribution considered in this work. We recall that in the WIP vertex both vector and axial components may participate in the spin-independent scattering, resulting in the following combinations that will be computed separately: vector-WIP vector-nucleus interaction, on the one hand, and axial-WIP vector-nucleus interaction, on the other hand; the former will be simply called vector scattering and the latter, axial scattering, both of them spin-independent. They are not to be confused with scalar interactions, namely scalar-WIP scalar-nucleus, which are also spin-independent. The vector and the axial currents can interfere with each other and are therefore considered together, whereas the scalar current can be treated separately. We will give details of the vector-axial decomposition of the WIP and nuclear currents in Sect.~\ref{dynamics}.
Our results focus on the above-mentioned vector and axial spin-independent scattering for a wide range of WIP energies and masses. Both processes have received relatively scarce attention in the literature. In the first case, vector scattering, the reason is that many typical dark matter candidates lack the vector current, being Majorana particles. This is the situation with most neutralino models. In the second case, axial scattering, the reason is that the corresponding cross sections are very small for typical dark matter kinematic conditions, namely large masses and low velocities. These circumstances will be discussed later. In this work, both processes become fully relevant since we consider a more general set of possible components of the dark matter content of the Universe, with wider ranges of mass and energy.

This paper is organized as follows. In Sect.~\ref{kinematics} we introduce the kinematics of the elastic scattering, and then we describe the dynamics in Sect.~\ref{dynamics}. Results are shown in Sect.~\ref{results}, and finally our main conclusions are given in Sect.~\ref{conclusions}.

\section{Kinematics}\label{kinematics}

Let $\vec{k}$ and $\varepsilon= \left(k^2+m^2\right)^{1/2}$ be the three-momentum and energy of the WIP before the scattering and $\vec{k}'$ and $\varepsilon'= \left(k'^2+m^2\right)^{1/2}$ the corresponding quantities after the scattering, where $m$ is the WIP mass. The incident momentum can be expressed in terms of the WIP mass and velocity as $k=\beta\:\gamma \:m$, where $\beta=v/c$ is the WIP velocity $v$ in units of the speed of light, and $\gamma=\left(1-\beta^2\right)^{-1/2}$. The energy transfer in the process is $\omega=\varepsilon-\varepsilon'$, and the momentum transfer is $q=\left(k^2+k'^2-2kk'\cos\theta\right)^{1/2}$, where $\theta$ is the angle between $\vec{k}$ and $\vec{k}'$, or scattering angle; in an elastic scattering the energy and momentum loss of the projectile target turns entirely into kinetic energy and momentum of the recoiling target, and consequently the energy and momentum transfers are related by 
\begin{eqnarray}
q=\left(2\:M\:\omega + \omega^2\right)^{1/2} \qquad \text{or} \qquad \omega = \sqrt{M^2 + q^2} - M \:,
\label{elastic_condition}
\end{eqnarray}
where $M$ is the target mass. The four-momentum transfer squared, $Q^2=\omega^2-q^2$, is related to the energy transfer in an elastic scattering through $|Q^2|=2M\omega$.
We define the dark matter limit (DML) as the kinematic scenario where the WIP mass is very large (possibly $m > M$) and its speed is small, keeping the incident momentum $k=\beta\:\gamma \:m$ small. These are features usually attributed to dark matter particles, that are generally known as weakly interacting {\it massive} particles, WIMPs. In this limit one has $\epsilon \approx m$ and $ \epsilon' \approx m$, so that $\omega\approx 0$. On the other side one has the extreme relativistic limit (ERL), where the WIP mass is very small (clearly $m << M$) and it moves at relativistic speed, again keeping the momentum small; in this situation $\epsilon\approx k$ and $\epsilon'\approx k'$. In both limits one has $\omega<<q$, which holds for elastic scattering in the small $q$ region, and more generally in the whole $q$ range as long as $k<<M$.

By fixing the incident WIP momentum $k$ or energy $\varepsilon$ and the scattering angle $\theta$, the momentum and energy of the outgoing WIP after the elastic scattering are given by:
\begin{eqnarray}
&& k'_{\pm} = \frac{1}{D} \:\left[ (M \:\varepsilon + m^2) \:(k\:\cos\theta) \pm k \:(M+\varepsilon) \:F^{1/2}  \right] \label{final_mom}\\
&& \varepsilon'_{\pm} = \frac{1}{D} \:\left[ (M \:\varepsilon + m^2) \:(M+\varepsilon) \pm k^2 \:\cos\theta \:F^{1/2}  \label{final_ener}\right] \:,
\end{eqnarray}
with
\begin{eqnarray}
&& D = (M + \varepsilon)^2 - k^2\:\cos^2\theta \;\; > 0 \\
&& F = M^2 - m^2\:\sin^2\theta \:.
\end{eqnarray}
In the DML with $m > M$ one must impose $F>0$, which implies that the scattering angle lies within the range $0\le\theta\le\theta_{max}$, with
\begin{eqnarray}
\theta_{max} = \arcsin(M/m) \le \pi/2 \;.
\end{eqnarray}
In this case both solutions $k'_+$ and $k'_-$ in Eq. (\ref{final_mom}), or $\varepsilon'_+$ and $\varepsilon'_-$ in Eq. (\ref{final_ener}), are valid. On the contrary, when $m < M$, as in the ERL, one has $F>0$ for every scattering angle, and the only physical values of the outgoing WIP momentum and energy are $k'_+$ and $\varepsilon'_+$, but not $k'_-$ and $\varepsilon'_-$.

\section{Dynamics}\label{dynamics}

The initial and final WIP spinors together with the interaction current give rise to the WIP four-tensor $\eta_{\mu\nu}$ that can be decomposed into a purely vector tensor ($VV$), a purely axial tensor ($AA$) and a vector-axial interference tensor ($VA$). Similarly, the nuclear spinors and the interaction current give rise to the nuclear four-tensor $W^{\mu\nu}$, that can be decomposed in the same manner. The vector and axial tensors are symmetric ($s$) whereas the interference tensors are antisymmetric ($a$) under exchange of indices. The WIP and nuclear tensors contract their symmetric and antisymmetric parts separately (cross-contractions vanish) to yield the WNC scattering matrix element squared $\widehat{{\mathcal R}}$ \cite{mor14}:
\begin{eqnarray}
\widehat{{\mathcal R}} &=& \frac{1}{v_0} \:\eta_{\mu\nu}\:W^{\mu\nu} = \frac{1}{v_0} \:\left[ \eta^s_{\mu\nu}\:W_s^{\mu\nu} + \eta^a_{\mu\nu}\:W_a^{\mu\nu} \right] = \frac{1}{v_0} \:\left[ \left( \eta^{VV}_{\mu\nu} + \eta^{AA}_{\mu\nu} \right)\:\left( W_{VV}^{\mu\nu} + W_{AA}^{\mu\nu} \right)+ \eta^{VA}_{\mu\nu}\:W_{VA}^{\mu\nu} \right] \:.
\end{eqnarray}
The contributions to the matrix element can then be regrouped as follows:
\begin{eqnarray}
\widehat{{\mathcal R}} = \frac{1}{v_0} \left( \eta^{VV}_{\mu\nu} + \eta^{AA}_{\mu\nu} \right)\:W_{VV}^{\mu\nu} + \frac{1}{v_0} \left( \eta^{VV}_{\mu\nu} + \eta^{AA}_{\mu\nu} \right)\:W_{AA}^{\mu\nu} + \frac{1}{v_0} \eta^{VA}_{\mu\nu}\:W_{VA}^{\mu\nu} = \widehat{{\mathcal R}}^V + \widehat{{\mathcal R}}^A + \widehat{{\mathcal R}}^{VA} \;,
\end{eqnarray}
where $\widehat{{\mathcal R}}^V$, $\widehat{{\mathcal R}}^A$, and $\widehat{{\mathcal R}}^{VA}$
are the contributions proportional to the nuclear tensors $W_{VV}$, $W_{AA}$, and $W_{VA}$, respectively. More specifically, each of them contains every possible multipole contribution ${\mathcal J}$ to the full matrix element:
\begin{eqnarray}
\widehat{{\mathcal R}} = \sum_{\mathcal J} \left(\widehat{{\mathcal R}}^V_{\mathcal J} + \widehat{{\mathcal R}}^A_{\mathcal J} + \widehat{{\mathcal R}}^{VA}_{\mathcal J} \right)\;,
\label{R}
\end{eqnarray}
which can be written in terms of generalized Rosenbluth factors (coming from the WIP four-tensor) and of nuclear form factors (from the nuclear target four-tensor) as:
\begin{eqnarray} \nonumber
\widehat{{\mathcal R}}^V_{\mathcal J} &=& \gamma_{V(0)}^2 \:\left[V_{L} \:\left(f^{V(0)}_{CC,\:{\mathcal J}}\right)^2 + V_{T} \:\left(f^{V(0)}_{T,\:{\mathcal J}}\right)^2 \right] +
\gamma_{V(1)}^2 \:\left[V_{L} \:\left(f^{V(1)}_{CC,\:{\mathcal J}}\right)^2 + V_{T} \:\left(f^{V(1)}_{T,\:{\mathcal J}}\right)^2 \right]  +\\
&&+ 2\:\gamma_{V(0)} \:\gamma_{V(1)} \:\left[V_{L} \:f^{V(0)}_{CC,\:{\mathcal J}} \:f^{V(1)}_{CC,\:{\mathcal J}} + V_{T} \:f^{V(0)}_{T,\:{\mathcal J}} \:f^{V(1)}_{T,\:{\mathcal J}} \right]  \label{RV} \;,\\\nonumber
\widehat{{\mathcal R}}^A_{\mathcal J} &=& \gamma_{A(0)}^2 \:\left[V_{LL} \:\left(f^{A(0)}_{LL,\:{\mathcal J}}\right)^2 + V_{T} \:\left(f^{A(0)}_{T,\:{\mathcal J}}\right)^2 \right] +
\gamma_{A(1)}^2 \:\left[V_{LL} \:\left(f^{A(1)}_{LL,\:{\mathcal J}}\right)^2 + V_{T} \:\left(f^{A(1)}_{T,\:{\mathcal J}}\right)^2 \right]  +\\
&&+ 2\:\gamma_{A(0)} \:\gamma_{A(1)} \:\left[V_{LL} \:f^{A(0)}_{LL,\:{\mathcal J}} \:f^{A(1)}_{LL,\:{\mathcal J}} + V_{T} \:f^{A(0)}_{T,\:{\mathcal J}} \:f^{A(1)}_{T,\:{\mathcal J}} \right]  \label{RA} \;,\\\nonumber
\widehat{{\mathcal R}}^{VA}_{\mathcal J} &=&  \gamma_{V(0)} \:\gamma_{A(0)} \:V_{T'} \:f^{V(0)}_{T,\:{\mathcal J}} \:f^{A(0)}_{T,\:{\mathcal J}} + \gamma_{V(0)} \:\gamma_{A(1)} \:V_{T'} \:f^{V(0)}_{T,\:{\mathcal J}} \:f^{A(1)}_{T,\:{\mathcal J}} +\\
&&+ \gamma_{V(1)} \:\gamma_{A(0)} \:V_{T'} \:f^{V(1)}_{T,\:{\mathcal J}} \:f^{A(0)}_{T,\:{\mathcal J}} + \gamma_{V(1)} \:\gamma_{A(1)} \:V_{T'} \:f^{V(1)}_{T,\:{\mathcal J}} \:f^{A(1)}_{T,\:{\mathcal J}} \;,
\label{RVA}
\end{eqnarray}
where $V$ and $A$ again stand for vector and axial, respectively; $C$, $L$ and $T$/$T'$ stand for Coulomb (time-like), longitudinal and transverse (spatial-like) components, respectively; and $(0)$ and $(1)$ stand for isoscalar and isovector components, respectively. The factors $\gamma_{V(0)}$, $\gamma_{V(1)}$, $\gamma_{A(0)}$, $\gamma_{A(1)}$ are nucleonic form factors, considered here just as coupling constants, whose values depend on the nature of the boson exchanged between the WIP and the hadronic target; if that boson is predominantly a $Z^0$, they would take values close to the SM weak neutral current (WNC) hadronic couplings: $\gamma_{V(0)}=\beta_{V(0)}=-2\sin^2\theta_W$, $\gamma_{V(1)}=\beta_{V(1)}=1-2\sin^2\theta_W$, $\gamma_{A(0)}=\beta_{A(0)}=0$, $\gamma_{A(1)}=\beta_{A(1)}=1$ (at tree level), where $\theta_W$ is the weak mixing angle ($\sin^2\theta_W\approx 0.23$). For the SM couplings we have used the notation and conventions in \cite{don79}. The nuclear (point-like) form factors $f$ themselves do not include couplings.

The generalized Rosenbluth factors $V$ in Eqs. (\ref{RV})-(\ref{RVA}) for elastic scattering can be written as \cite{mor14}:
\begin{eqnarray}
&& V_{L} = b_V^2 \:\frac{1}{2}\left[ 1-2\nu^2+\nu^4\right] + b_A^2
\:\frac{1}{2}\left[ 1-\frac{4\:m^2}{v_0} -\left(2-\frac{4\:m^2}{v_0}\right)\:\nu^2+\nu^4\right] \;, \label{Rosenbluth_VL}\\
&& V_{LL} = b_V^2 \:\frac{1}{2}\:\nu^2 + b_A^2 \:\frac{1}{2}\left[ \frac{4\:m^2}{v_0} + \nu^2 \right] \;, \label{Rosenbluth_VLL}\\
&& V_{T} = b_V^2 \:\frac{1}{4}\left[ 1- \frac{4\:m^2}{v_0} + \frac{4\:M\:\omega}{v_0} - \nu^2 \right] + b_A^2 \:\frac{1}{4}\left[ 1+ \frac{4\:m^2}{v_0} + \frac{4\:M\:\omega}{v_0} - \nu^2 \right] \;, \label{Rosenbluth_VT}\\
&& V_{T'} = b_V \:b_A \:\frac{2\:M\:(\epsilon+\epsilon')}{v_0}\:\nu \;, \label{Rosenbluth_VTp}
\end{eqnarray}
where $\nu=\omega/q$ and $v_0=(\epsilon+\epsilon')^2-q^2$. The effective vector and axial WIP couplings to the exchanged boson(s) are respectively $b_V$ and $b_A$. For SM neutrinos, exchanging a $Z^0$, one has \mbox{$b_V=a^{\nu}_V=1$} and \mbox{$b_A=a^{\nu}_A=1$}.

In the DML one has $4\:m^2/v_0 \to 1$, and using the fact that $\nu \to 0$, the Rosenbluth factors in Eqs. (\ref{Rosenbluth_VL})-(\ref{Rosenbluth_VTp}) become
\begin{eqnarray}
&& V^{DML}_{L} = \frac{1}{2} \:b_V^2 \;,\\
&& V^{DML}_{LL} = \frac{1}{2} \:b_A^2 \;,\\
&& V^{DML}_{T} = \frac{1}{2} \:b_A^2 \;,\\
&& V^{DML}_{T'} = 0 \;.
\label{Rosenbluth_DML}
\end{eqnarray} 
The time component of the DML axial current vanishes in this limit, as can be seen in $V^{DML}_L$, whereas the spatial components of the DML vector current vanish, as seen in $V^{DML}_{LL}$ and in $V^{DML}_{T}$. Therefore, for Majorana particles, where the vector current vanishes \cite{vog98}, only the space-like Rosenbluth factors are involved. On the hadronic vertex, the space components come predominantly from the axial current (see tables with relative sizes of multipole operators in \cite{don79} or \cite{mor15}); as a consequence, for Majorana particles the dominant interaction is axial-axial, or spin-dependent (see also the discussion in \cite{eng92}).

In the ERL one has $4\:m^2/v_0 \to 0$, and using again $\nu \to 0$, the Rosenbluth factors are:
\begin{eqnarray}
&& V^{ERL}_{L} = \frac{1}{2} \:(b_V^2+b_A^2) \;,\\
&& V^{ERL}_{LL} = 0\;,\\
&& V^{ERL}_{T} = \frac{1}{4} \:(b_V^2+b_A^2) \left[\frac{1}{2}+\tan^2\theta/2\right] \;,\\
&& V^{ERL}_{T'} = b_V \:b_A \:\frac{\sin\theta/2}{\cos^2\theta/2} \;.
\label{Rosenbluth_ERL}
\end{eqnarray} 

In addition to the matrix element of the vector current vanishing for Majorana particles, effectively having $b^{(M)}_V=0$, the Majorana matrix element of the axial current is twice as large as for Dirac particles, effectively having $b^{(M)}_A=2\:b_A$, both results coming from the fact that the charge conjugate of a Majorana fermion field differs from the field itself just by a phase \cite{vog98}. In what follows we will keep the results for Dirac particles and assume that effective couplings $b^{(M)}_{V,A}$ as described above are to be used for Majorana particles.

We now discuss the coherent contribution to the elastic scattering cross section, where all the nucleons participate coherently and is therefore the most important for nuclei heavier than $^3$He; it depends on the Coulomb vector isoscalar monopole form factor,
\begin{eqnarray}
\widehat{{\mathcal R}}_{coh} = \gamma_{V(0)}^2 \:V_{L} \:\left(f^{V(0)}_{CC,\:{\mathcal J}=0}\right)^2 \;.
\label{R_coh}
\end{eqnarray} 
Full coherence applies for momentum transfers corresponding to nuclear-size wavelengths and below, \mbox{$q\lesssim 160 \:A^{-1/3}$ MeV}; for larger values the Coulomb form factor decreases and the incoherent elastic form factors may become comparable. For spin-0 ($J=0$), isospin-0 ($T=0$) nuclear targets, namely even-even nuclei with $N=Z$ and neglecting isospin mixing, the coherent contribution is actually the only one possible in elastic scattering; with non-pure 0 isospin targets ($T\ne 0$) there is an additional, still Coulomb-like, contribution, and for other nuclei ($J\ne 0$) the general expression in Eq. (\ref{R}) applies, but in all cases the coherent contribution is dominant except for very light nuclei.
In the DML this predominant coherent matrix element squared is simply given by
\begin{eqnarray}
\widehat{{\mathcal R}}^{DML}_{coh} = \frac{1}{8} \:b_V^2 \:\gamma_{V(0)}^2 \:A^2 \;,
\label{R_coh_DML}
\end{eqnarray} 
whereas in the ERL one has
\begin{eqnarray}
\widehat{{\mathcal R}}^{ERL}_{coh} = \frac{1}{8} \:\left(b_V^2+b_A^2\right) \:\gamma_{V(0)}^2 \:A^2 \;.
\label{R_coh_ERL}
\end{eqnarray} 
In both cases the coherence of the process is manifest through the dependence on the number of nucleons squared, $A^2$.

From the squared matrix elements discussed until now one can compute the cross section of the elastic WIP-nucleus scattering as
\begin{eqnarray}\label{xs}
\left( \frac{d\sigma}{d\Omega} \right) = \frac{\kappa^2}{8\pi^2} \:\frac{k'}{k} \:F_{rec}^{-1} \:v_0 \:\widehat{{\mathcal R}} \:,
\end{eqnarray}
which is valid when a massive boson is exchanged, so that its mass is much larger than the carried four-momentum, $M_B >> Q$,  and where $\kappa$ is an effective overall coupling constant inversely proportional to $M_B^2$; for $Z^0$ exchange, namely the SM WNC, $\kappa = G_F$. This cross section is valid for a fully polarized beam, as the one for SM neutrinos; for other situations, a polarization factor ${\mathcal P}$ lower than 1 should be included. The recoil factor in Eq. (\ref{xs}) is given by
\begin{eqnarray}\label{recoil}
F_{rec} = \left| 1 + \frac{\varepsilon \:k'-\varepsilon' \:k}{k' \:M} \right| \;.
\end{eqnarray}
The coherent cross section can be given separately for the vector and for the axial components using Eqs. (\ref{R_coh}) and (\ref{Rosenbluth_VL}), as
\begin{eqnarray}
\left( \frac{d\sigma}{d\Omega} \right)^{coh(V)} &=& \frac{\kappa^2}{16\pi^2} \:b_V^2 \:\gamma_{V(0)}^2 \:\frac{k'}{k} \:F_{rec}^{-1} \:v_0 \:\left[ 1-2\nu^2+\nu^4\right] \:\left(f^{VV(0)}_{CC} \right)^2 \label{xs_coh_V}\\
\left( \frac{d\sigma}{d\Omega} \right)^{coh(A)} &=& \frac{\kappa^2}{16\pi^2} \:b_A^2 \:\gamma_{V(0)}^2 \:\frac{k'}{k} \:F_{rec}^{-1} \:v_0 \:\left[ 1- \frac{4m^2}{v_0} -\left(2 - \frac{4m^2}{v_0} \right)\:\nu^2+\nu^4\right] \:\left(f^{VV(0)}_{CC} \right)^2 \;.
\label{xs_coh_A}
\end{eqnarray}
The ratio of the axial to the vector cross sections can then be written as:
\begin{eqnarray}
\frac{\left( \frac{d\sigma}{d\Omega} \right)^{coh(A)} }{\left( \frac{d\sigma}{d\Omega} \right)^{coh(V)}} &=& \frac{b_A^2}{b_V^2} \left[1-\frac{4m^2}{v_0} \left( \frac{1-\nu^2}{1-2\nu^2+\nu^4} \right)\right] \to \frac{b_A^2}{b_V^2} \left[1-\frac{4m^2}{v_0} \right] \;,
\label{ratio_xs}
\end{eqnarray}
where the limit refers to $\nu\to 0$. This ratio evaluated at small momentum transfer also provides a good estimation of the ratio of axial-to-vector integrated cross sections, $\sigma^{coh(A)}/\sigma^{coh(V)}$, since small $q$ values give the largest contribution to the integrated cross sections; it tends to 0 in the DML and to $b_A^2/b_V^2$ in the ERL.

\vskip1cm
We can establish a useful relationship between the WIP-nucleus and the electron-nucleus coherent cross sections through the parity-violating (PV) asymmetry in elastic electron-nucleus scattering \cite{mor15}:
\begin{eqnarray}\label{rel_massive}
&&\left( \frac{d\sigma}{d\Omega} \right)^{coh}\hspace{-0.4cm}(k,\bar{\theta}) = \frac{\kappa^2\:\gamma^2_{V(0)}}{G_F^2\:\beta^2_{V(0)}} \quad{\mathcal K}(k,k_e,\theta_e) \quad\mathcal{A}_{(e,e)}^2(\widehat{k}_e,\widehat{\theta}_e) \quad \left( \frac{d\sigma}{d\Omega} \right)^{coh}_{(e,e)}\hspace{-0.6cm}(k_e,\theta_e) \:,
\end{eqnarray}
where $\mathcal{A}_{(e,e)}$ is the PV asymmetry, defined as the relative difference between the cross sections of electrons with spin projection parallel (same direction, helicity $h=+1$) and antiparallel (opposite direction, helicity $h=-1$) to their momentum:
\begin{equation}\label{asymmetry_sigmas}
\mathcal{A}_{(e,e)} = \frac{\left(\frac{d\sigma}{d\Omega}\right)^{h=+1} - \left(\frac{d\sigma}{d\Omega}\right)^{h=-1}}{\left(\frac{d\sigma}{d\Omega}\right)^{h=+1} + \left(\frac{d\sigma}{d\Omega}\right)^{h=-1}} \;.
\end{equation}
The interest of this relationship lies in the fact that the electron-nucleus cross section can be measured easily, whereas significant progress is being made for the measurement of the PV asymmetry in elastic electron scattering (as in the PREX or CREX experiments), and both results can then be connected to the cross section of a weak interacting particle. In Eq. (\ref{rel_massive}) the kinematic dependences of each quantity are explicitly shown. As before, $\kappa\:\gamma_{V(0)}$ is the full hadronic vector isoscalar coupling to the exchanged particle, and $G_F\:\beta_{V(0)}$ is the full hadronic vector isoscalar SM WNC coupling; when the WIP and the hadronic target exchange a $Z^0$, $\kappa\:\gamma_{V(0)}=G_F\:\beta_{V(0)}$. The relationship in Eq. (\ref{rel_massive}) is strictly valid within the plane-wave Born approximation (PWBA), {\it i.e.}, neglecting the effect of the nuclear Coulomb field on the scattered electron wave function; measurements of the electron cross section and PV asymmetry in the left-hand side of the equation naturally contain the Coulomb distortion effect, but it can easily be taken into account. The kinematic factor ${\mathcal K}$ depends on the electron incident momentum $k_e$ and the scattering angle $\theta_e$ (the latter through the energy transfer $\omega_{e}$) and on a chosen WIP incident momentum $k$ (or energy $\varepsilon$), as well as on the WIP and the target masses, $m$ and $M$:
\begin{widetext}
\begin{eqnarray}\label{rel_mass_erl}
{\mathcal K}(k,k_e,\theta_e) &=&
\frac{ k^2_{e} \:\left[ k^2+w_{e}\:(-2\:\varepsilon +w_{e}) \right]^{3/2} }
{k \:\left[ k_{e}-w_{e} \right]^2 \:\left[ 2\:k_{e}^2-w_{e}\:(2\:k_{e}+M) \right] \:\left[k^2-w_{e}\:(\varepsilon +m^2) \right]} \times \\\nonumber
&\times & \frac{1}{2\:(a_A^{e})^2} \:\left\{ (b_V)^2 \:\left[2\:\varepsilon ^2-\omega_{e}\:(2\:\varepsilon +M) \right]
+(b_A)^2 \:\left[2\:k^2-\omega_{e}\:(2\:\varepsilon +M+m^2/M) \right] \right\} \;,
\end{eqnarray}
\end{widetext}
where $a_A^e$ is the electron axial WNC coupling ($a_A^e=1$ in the SM at tree level) and $b_V$, $b_A$ are the WIP couplings. The energy transfer (nuclear recoil energy) is
\begin{eqnarray}
\omega_{e} = \frac{2\:k^2_{e}\:\sin^2(\theta_{e}/2)}{M+2\:k_{e}\:\sin^2(\theta_{e}/2)} \;.
\end{eqnarray}

In Eq. (\ref{rel_massive}) the scattering angle $\bar{\theta}$ for which the WIP cross section is obtained can be computed as
\begin{eqnarray}
\bar{\theta} = \arccos\left[\frac{k^2-\omega_{e}\:(\varepsilon+M)}{k\:[k^2+\omega_{e}\:(\omega_{e}-2\:\varepsilon)]^{1/2}}\right] \;,
\label{angle}
\end{eqnarray}
and fulfills the condition $|\sin\bar{\theta}| \le M/m$.

\section{Results}\label{results}

In what follows we show results for WIP-nucleus elastic scattering cross sections using a light target, $^{12}$C ($Z=N=$ 6), and a heavy target, $^{208}$Pb ($Z=$ 82, $N=$ 126), and for a variety of WIP masses and velocities. The main difference between the $^{12}$C and the $^{208}$Pb targets is that the elastic scattering cross section of the latter is roughly 300 times larger than the former, as corresponds to the process being driven by the coherent contribution, proportional to $(N+Z)^2$. In addition, the coherent cross section for the $^{12}$C target is approximately equal to the full elastic cross section, since the isovector contribution is negligible (only due to small isospin mixing); for $^{208}$Pb, on the contrary, although the coherent contribution is as expected dominant, there is an additional isovector contribution to the elastic cross section, which is actually the largest among the stable nuclei ($N/Z$ is 1.54 for $^{208}$Pb).

\begin{figure}
\begin{center}
\includegraphics[trim= 0.5cm 1cm 2cm 1cm, clip, width=0.48\textwidth]{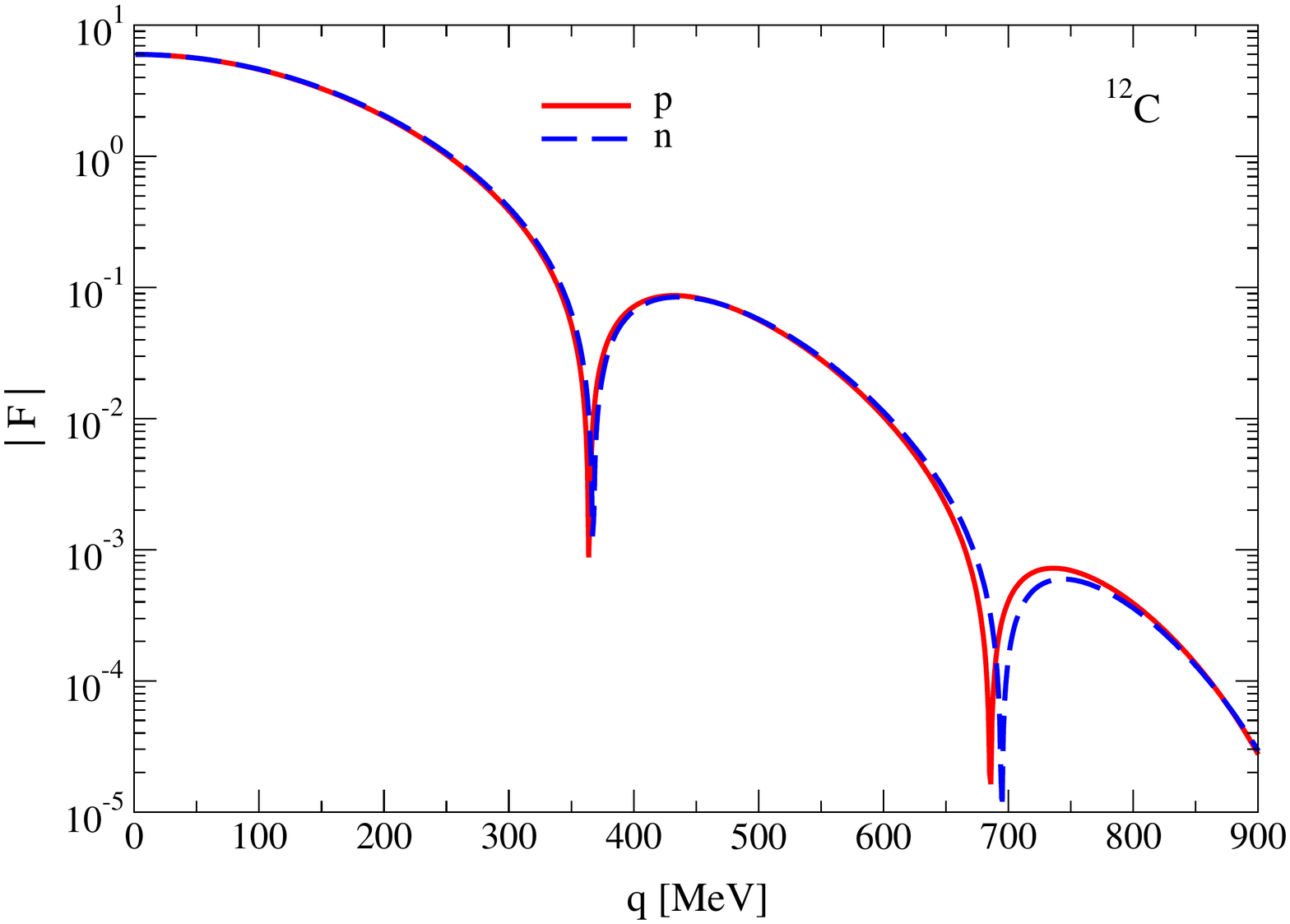}\hspace{0.3cm}
\includegraphics[trim= 0.5cm 1cm 2cm 1cm, clip, width=0.48\textwidth]{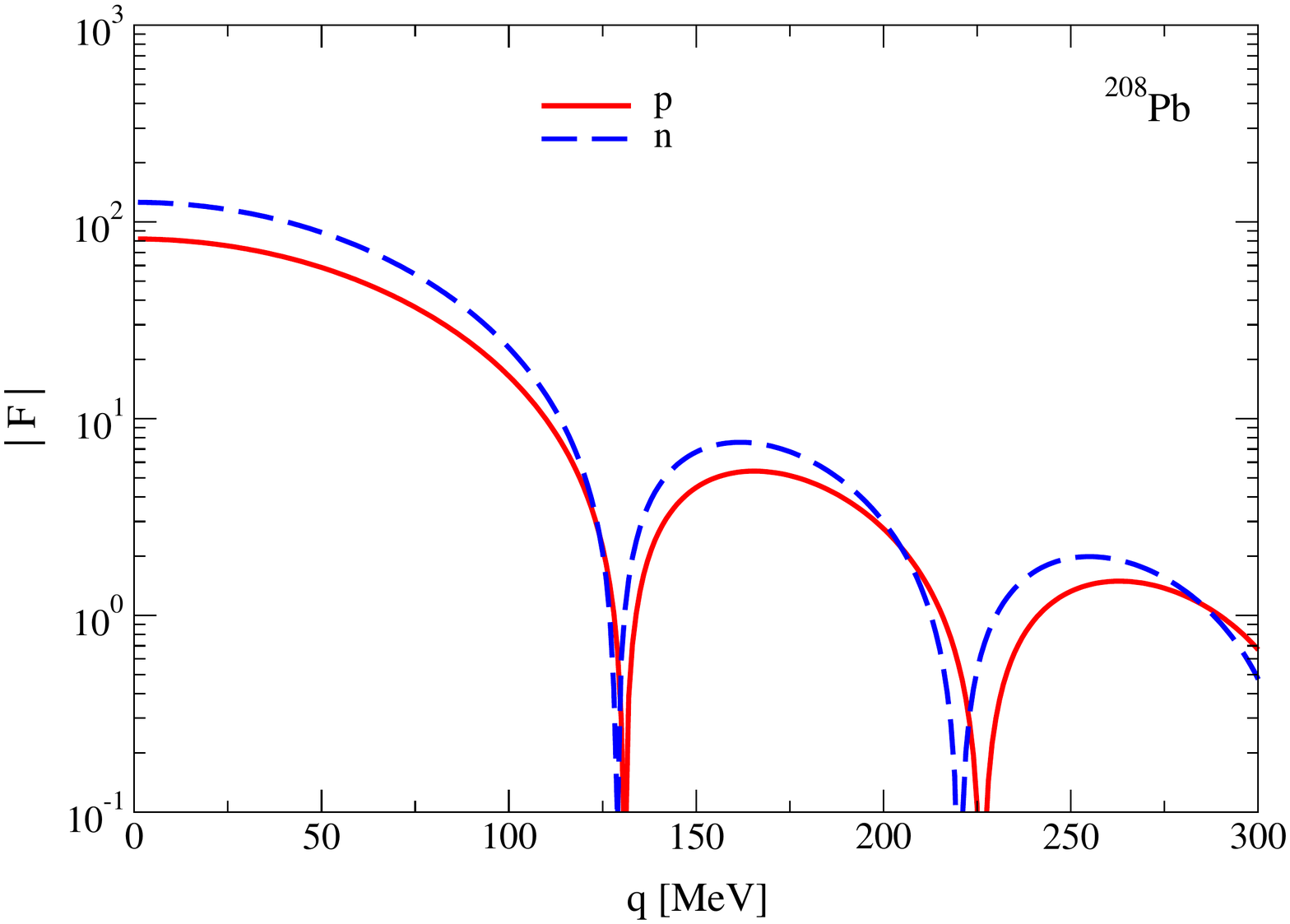}
\caption{(Color online) Proton and neutron distribution form factors as a function of the momentum transfer $q$ of the nuclear targets: $^{12}$C (left panel) and $^{208}$Pb (right panel).
\label{formfactors}}
\end{center}
\end{figure}

\begin{figure}
\begin{center}
\includegraphics[trim= 0cm 1cm 2cm 1cm, clip, width=0.48\textwidth]{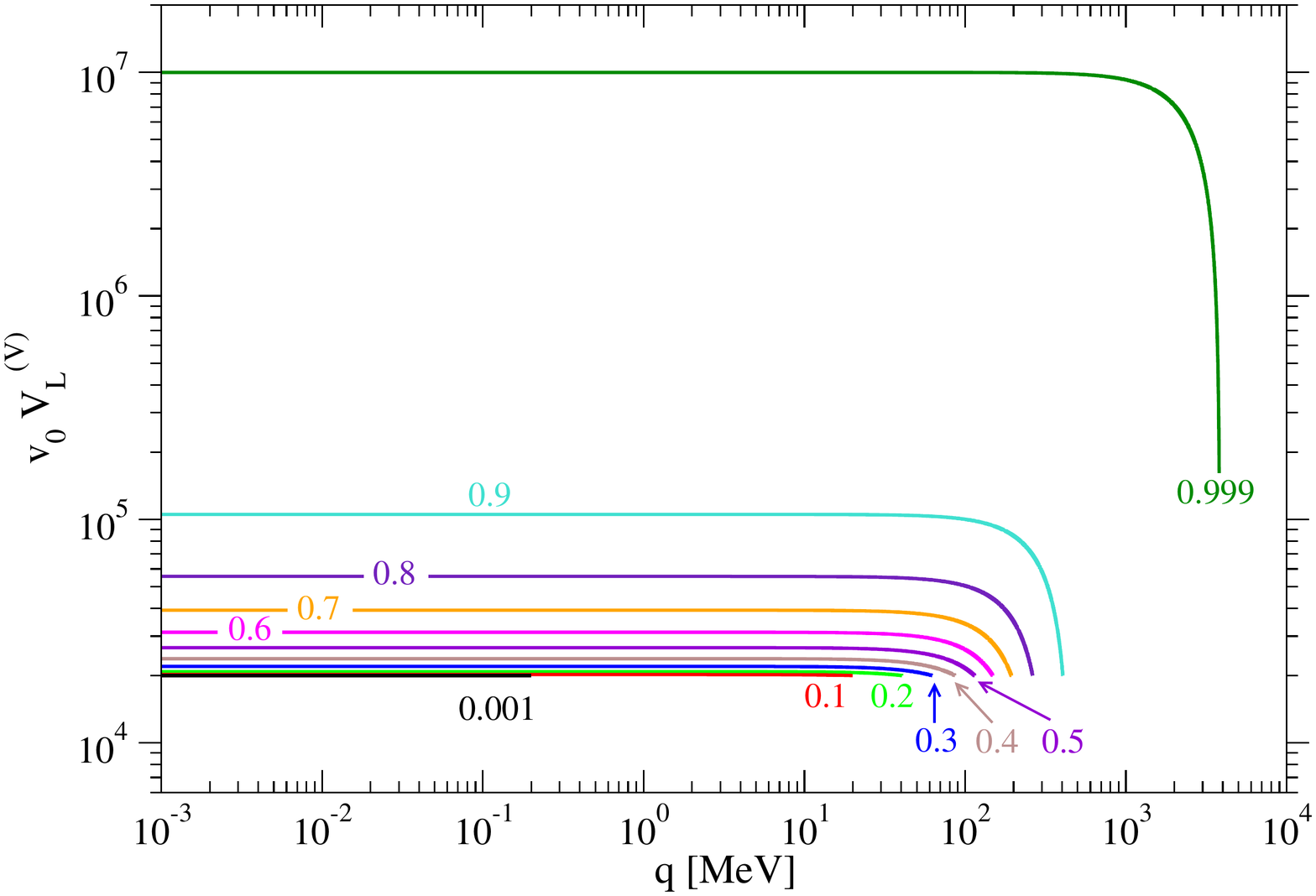}\hspace{0.3cm}
\includegraphics[trim= 0cm 1cm 2cm 1cm, clip, width=0.48\textwidth]{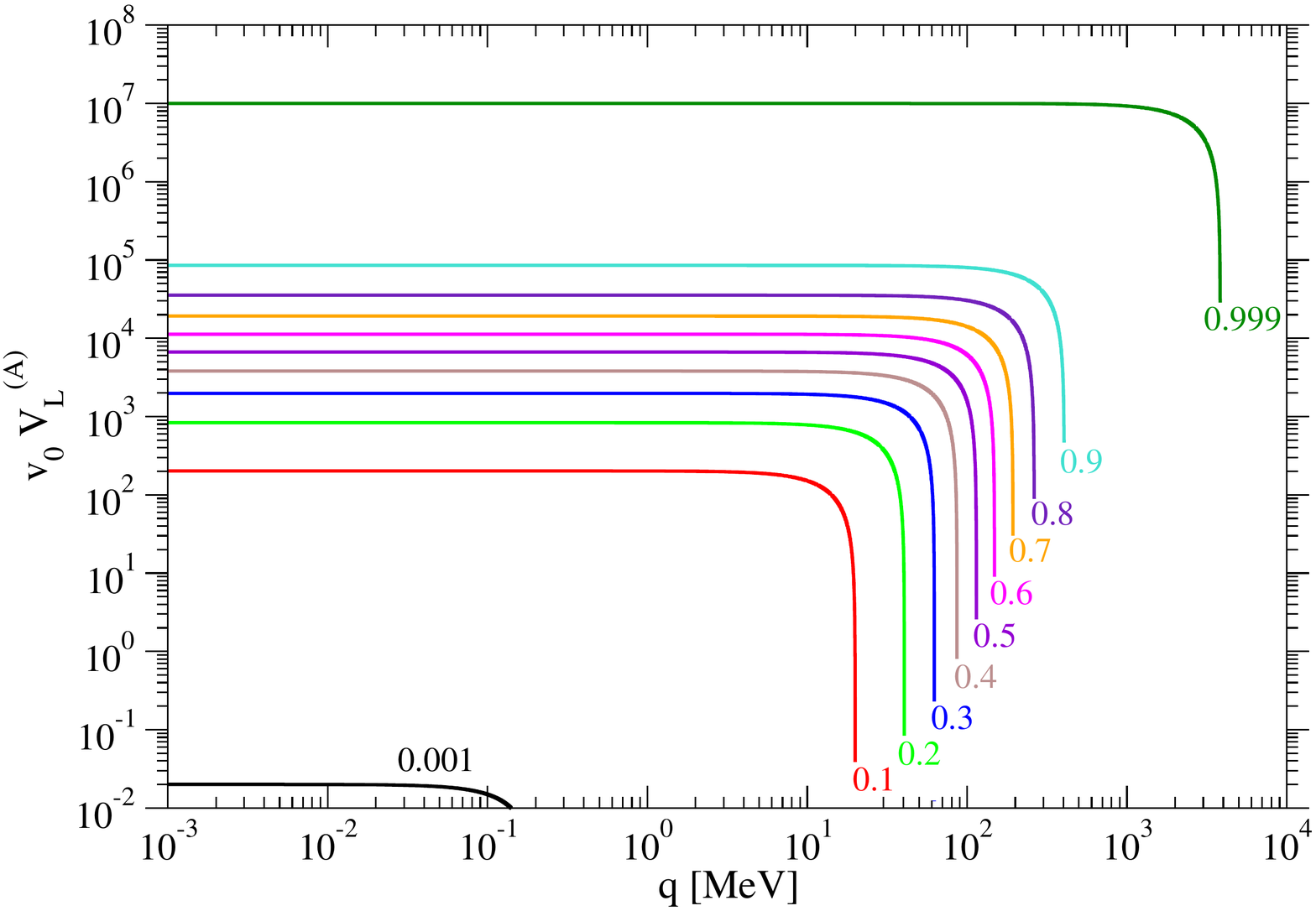}
\caption{(Color online) Longitudinal Rosenbluth factors including the factor $v_0$ as a function of the momentum transfer $q$ for a WIP mass of 100 MeV and different velocity parameters $\beta$ (given in the figure) for scattering from $^{12}$C: vector, $v_0V_L^{(V)}$ (left panel) and axial, $v_0V_L^{(A)}$ (right panel).
\label{rosenbluth}}
\end{center}
\end{figure}

We start by showing in Fig. \ref{formfactors} the proton and neutron distribution form factors of $^{12}$C and $^{208}$Pb as a function of the momentum transfer.  They have been obtained using an axially symmetric Skyrme Hartree-Fock mean field with BCS pairing for the ground state structure \cite{sar89}. In Fig. \ref{rosenbluth} we show the vector (to the left) and axial (to the right) longitudinal Rosenbluth factors for $^{12}$C for a WIP mass of 100 MeV and several velocities; results are very similar for $^{208}$Pb. The results of both Fig. \ref{formfactors} and Fig. \ref{rosenbluth} are the basic ingredients of the cross sections to be shown in the following figures.

\begin{figure}
\begin{center}
\includegraphics[trim= 0cm 1cm 2cm 1cm, clip, width=0.48\textwidth]{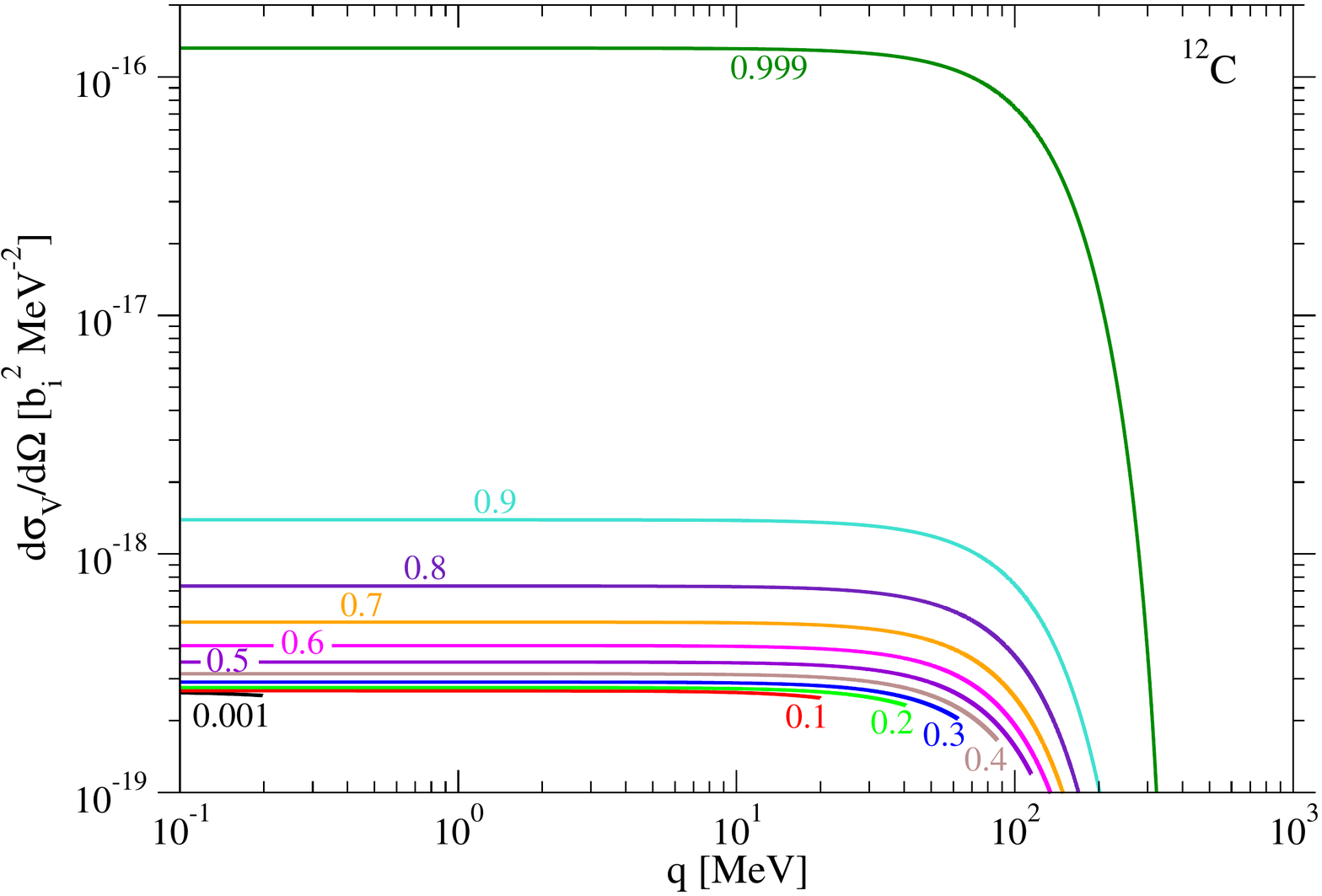}\hspace{0.3cm}
\includegraphics[trim= 0cm 1cm 2cm 1cm, clip, width=0.48\textwidth]{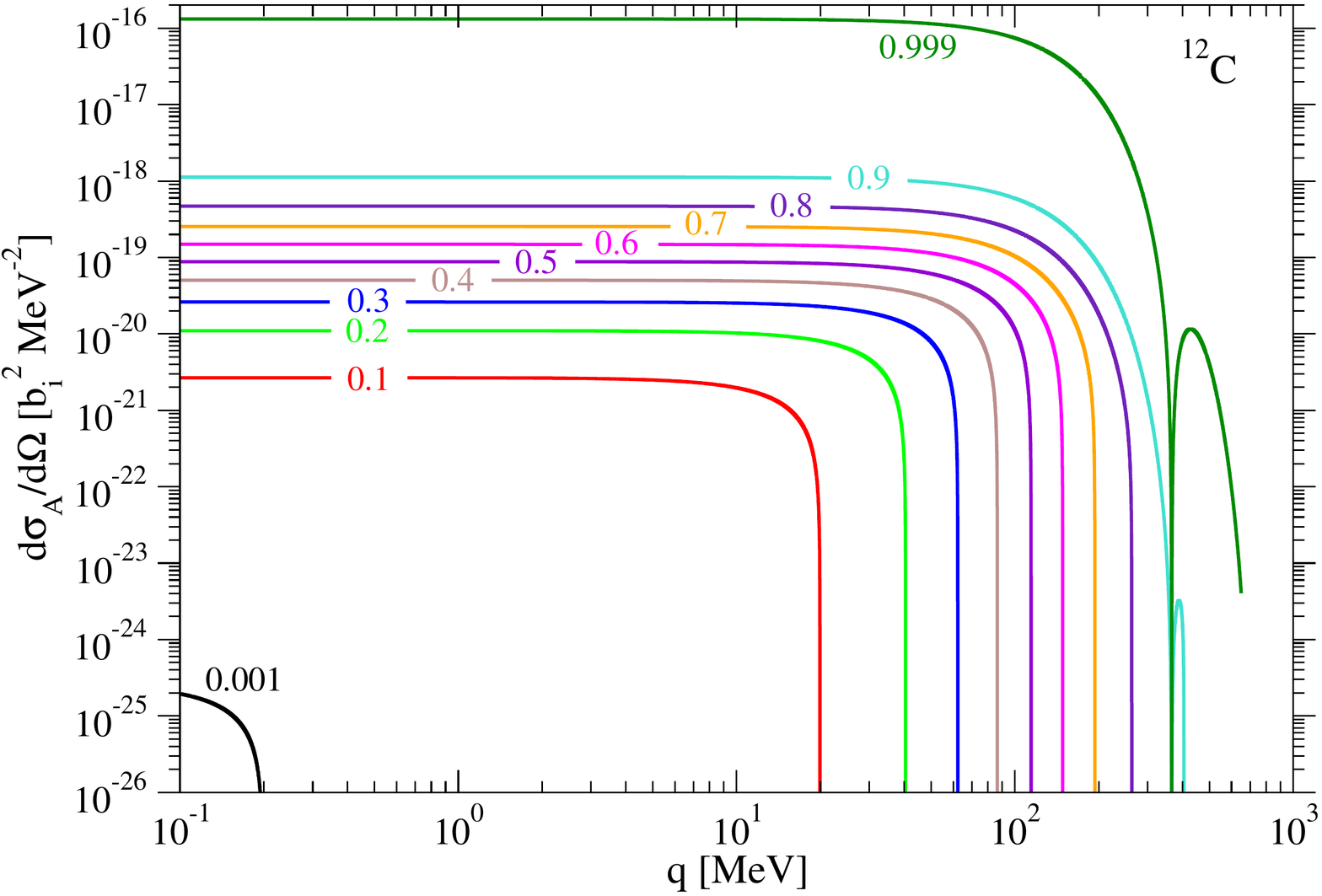}
\caption{(Color online) Differential cross section as a function of the momentum transfer $q$ for a WIP mass of 100 MeV and different velocity parameters $\beta$ (given in the figure) for scattering from $^{12}$C: vector contribution (left panel) and axial contribution (right panel).
\label{diff_xs_c12}}
\end{center}
\end{figure}

\begin{figure}
\begin{center}
\includegraphics[trim= 0cm 1cm 2cm 1cm, clip, width=0.48\textwidth]{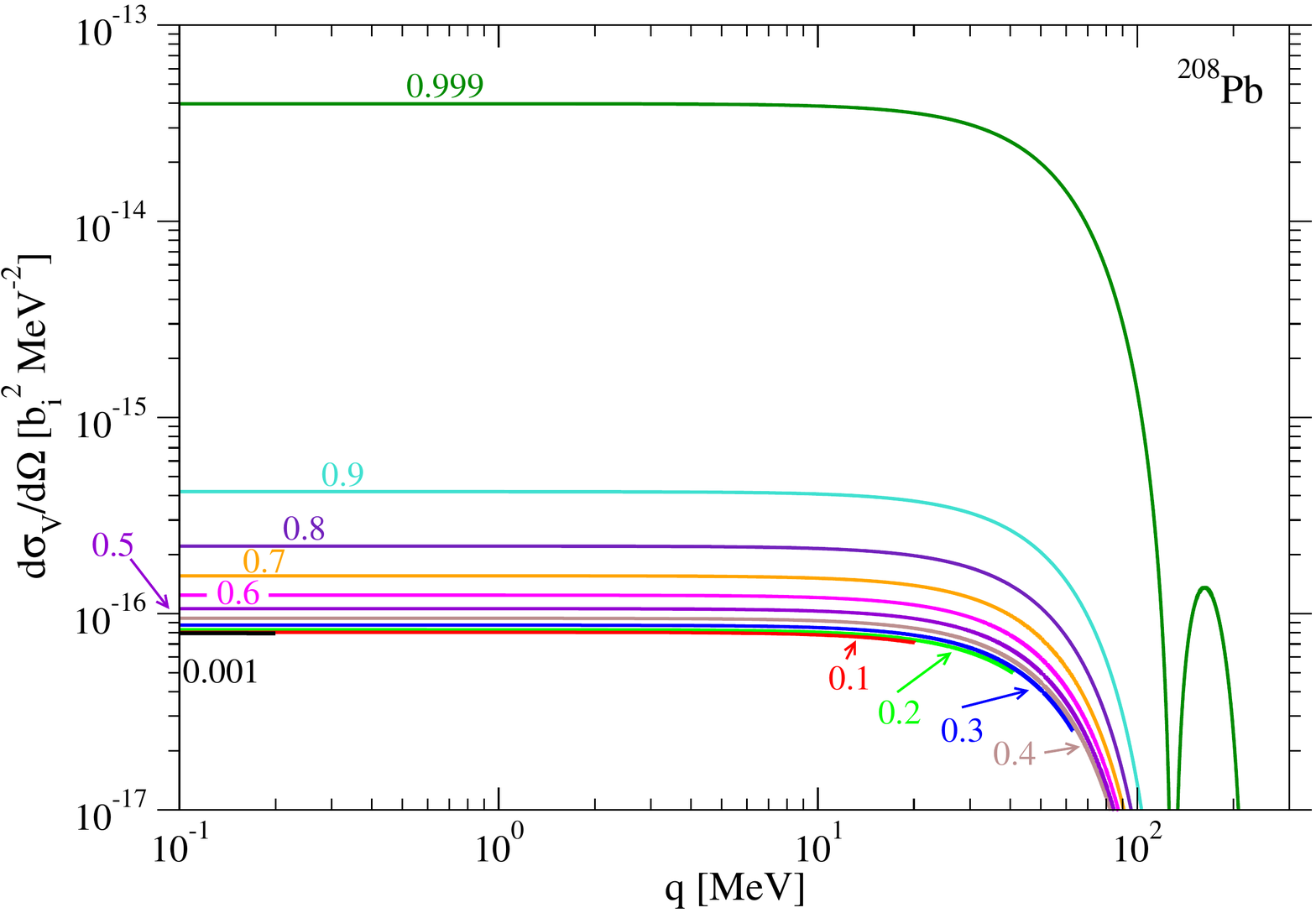}\hspace{0.3cm}
\includegraphics[trim= 0cm 1cm 2cm 1cm, clip, width=0.48\textwidth]{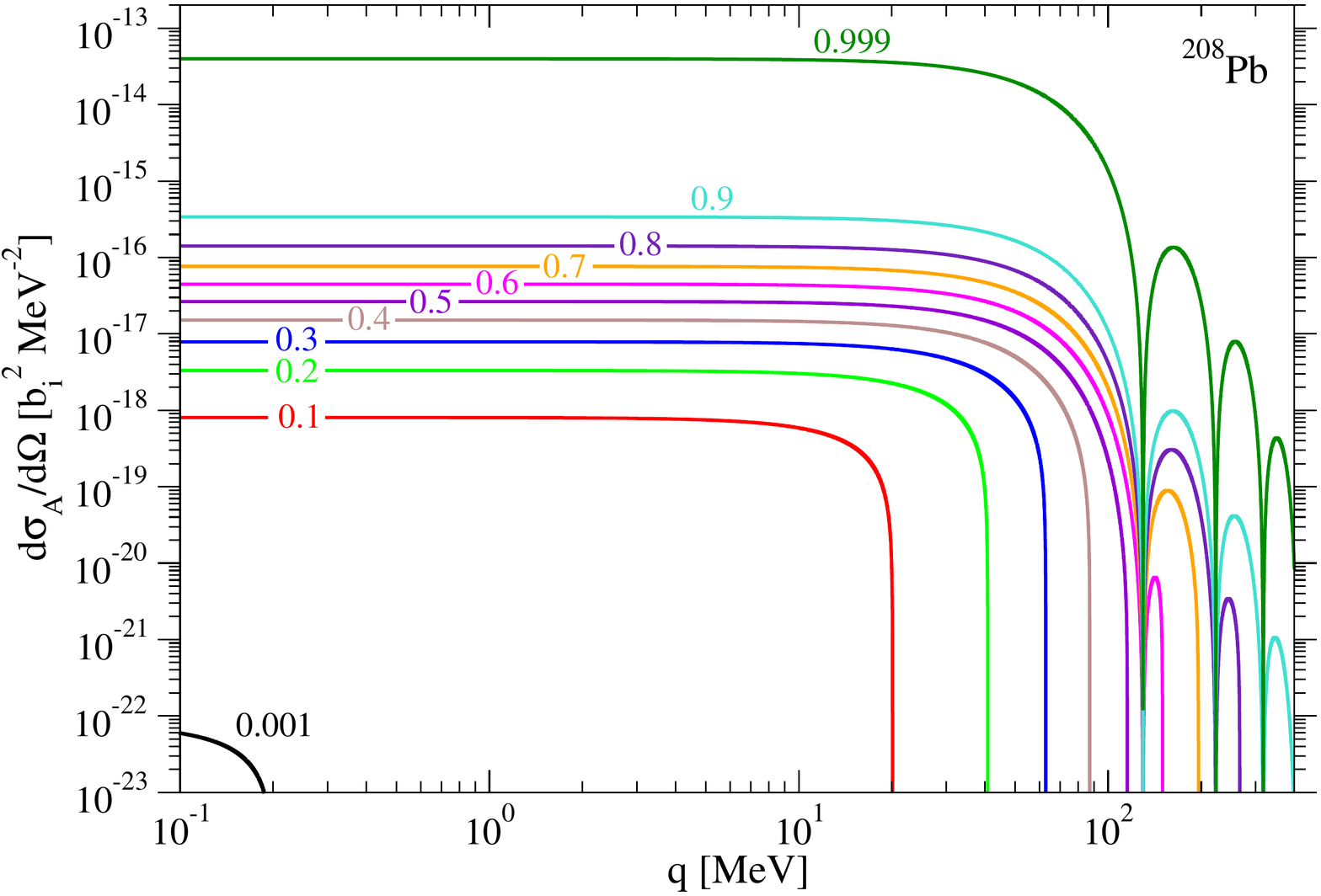}
\caption{(Color online) As for Fig. \ref{diff_xs_c12}, but now for a $^{208}$Pb target.
\label{diff_xs_pb208}}
\end{center}
\end{figure}

In Fig. \ref{diff_xs_c12} we show the vector (to the left) and axial (to the right) contributions to the differential WIP-$^{12}$C coherent cross section, Eqs. (\ref{xs_coh_V}) and (\ref{xs_coh_A}), as a function of the momentum transfer, for different WIP velocities $\beta=v/c$. The same results are shown in Fig. \ref{diff_xs_pb208} but for a $^{208}$Pb target. The cross sections are given in units of the WIP couplings $b_V^2$ (vector contributions) or $b_A^2$ (axial contributions), but the WNC value has been used for the overall and hadronic couplings, $G_F^2\:\beta_{V(0)}^2$, {\it i.e.}, the exchange of a $Z^0$ has been considered for the sake of example. For a different intermediate boson, these results should be multiplied by $\kappa^2\:\gamma_{V(0)}^2/(G_F^2\:\beta_{V(0)}^2)$, and in order to obtain the full value of the cross section, by the WIP couplings $b_V^2$ for the vector and $b_A^2$ for the axial contributions, respectively.

\begin{figure}
\begin{center}
\includegraphics[trim= 0cm 0cm 0cm 0cm, clip, width=0.48\textwidth]{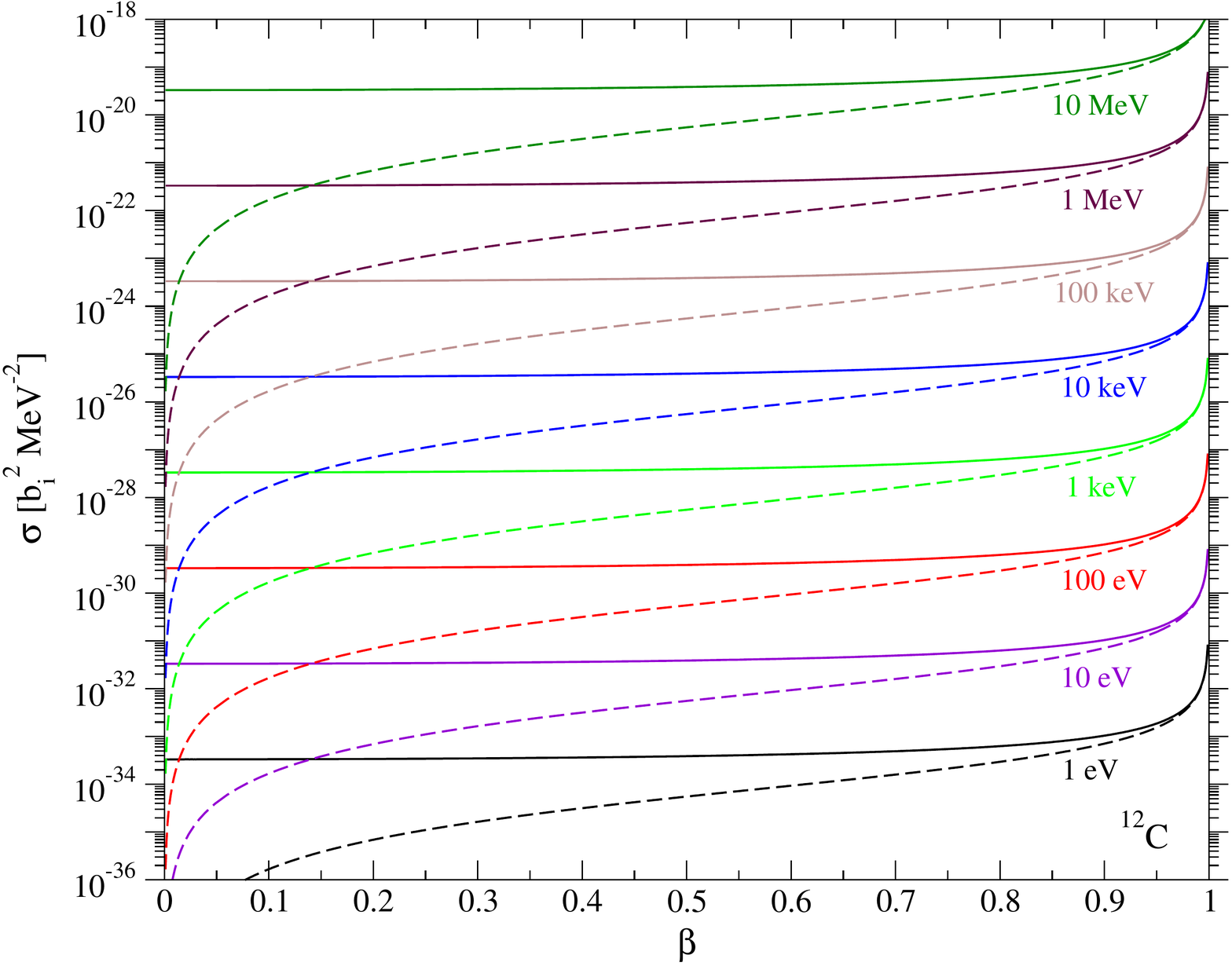}\hspace{0.5cm}
\includegraphics[trim= 0cm 0cm 0cm 0cm, clip, width=0.48\textwidth]{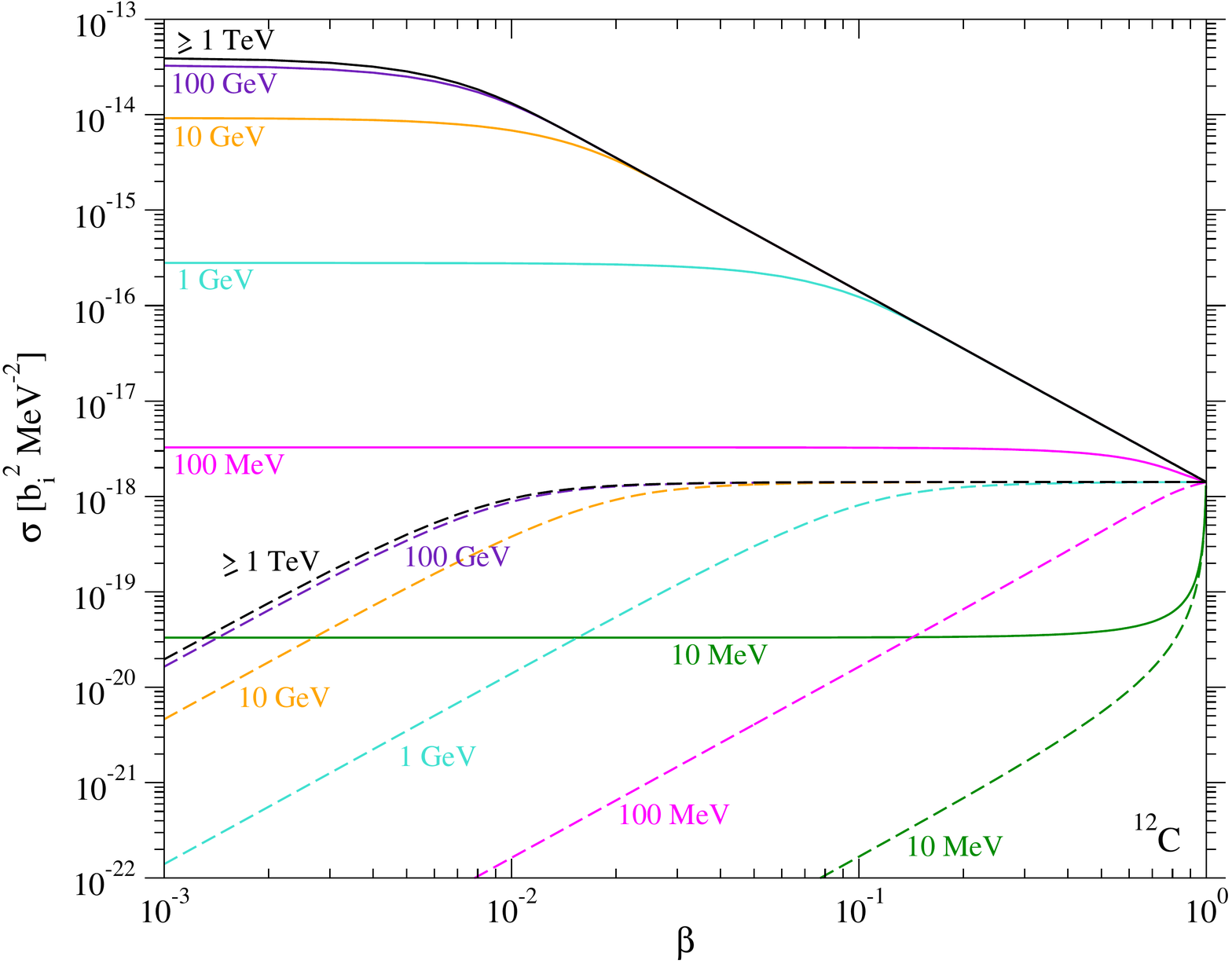}
\caption{(Color online) Vector (solid lines, in units $b^2_V$ MeV$^{-2}$) and axial (dashed lines, in units $b^2_A$ MeV$^{-2}$) contributions to the integrated cross sections of the WIP-$^{12}$C interaction through $Z^0$ exchange, as a function of the WIP velocity in terms of $\beta=v/c$ for different WIP masses. Left panel: WIP masses equal or below 10 MeV, with a linear abscissa. Right panel: WIP masses equal or above 10 MeV, with a logarithmic abscissa.
\label{xs_vs_vel_c12}}
\end{center}
\end{figure}

\begin{figure}
\begin{center}
\includegraphics[trim= 0cm 0cm 0cm 0cm, clip, width=0.48\textwidth]{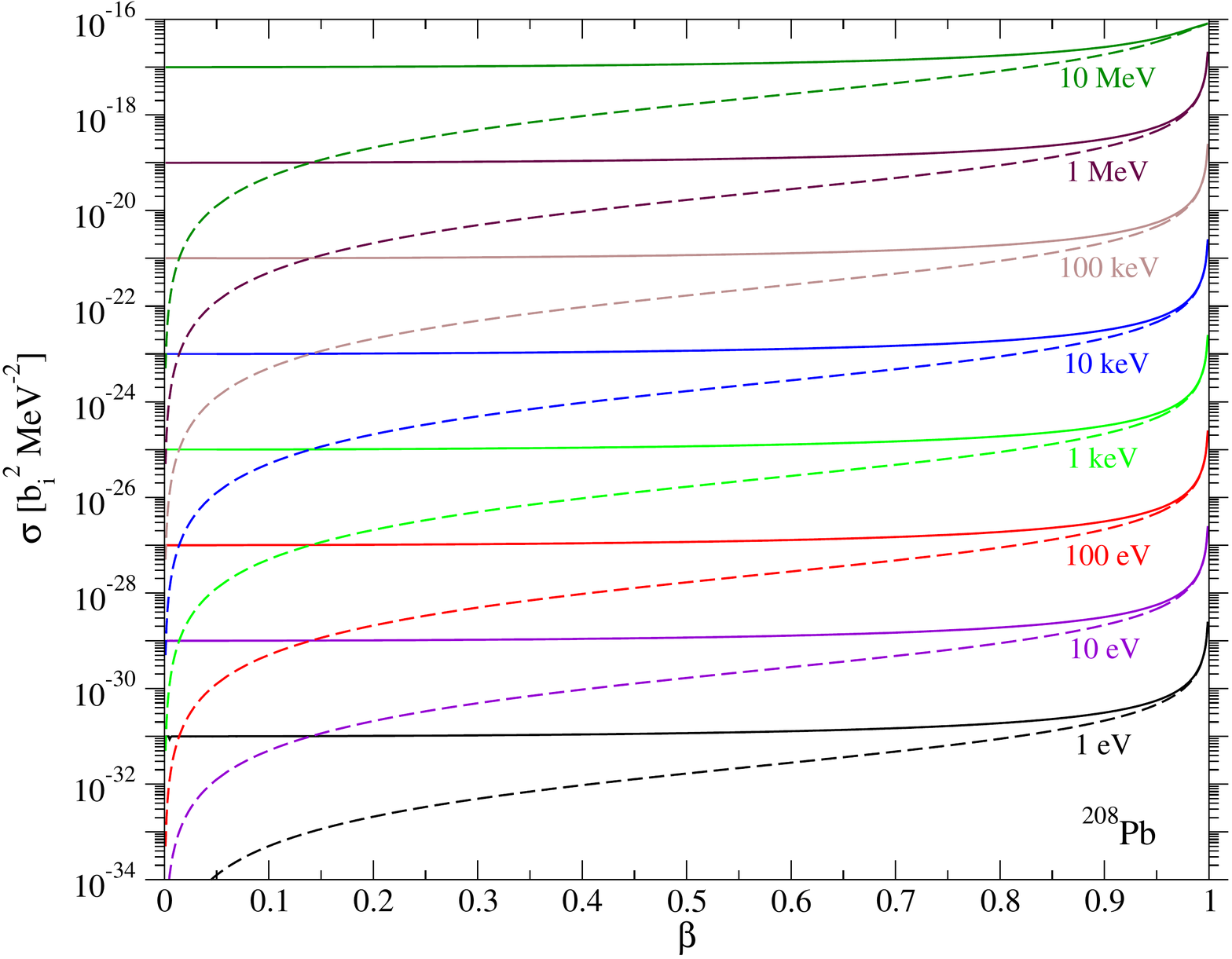}
\hspace{0.5cm}
\includegraphics[trim= 0cm 0cm 0cm 0cm, clip, width=0.48\textwidth]{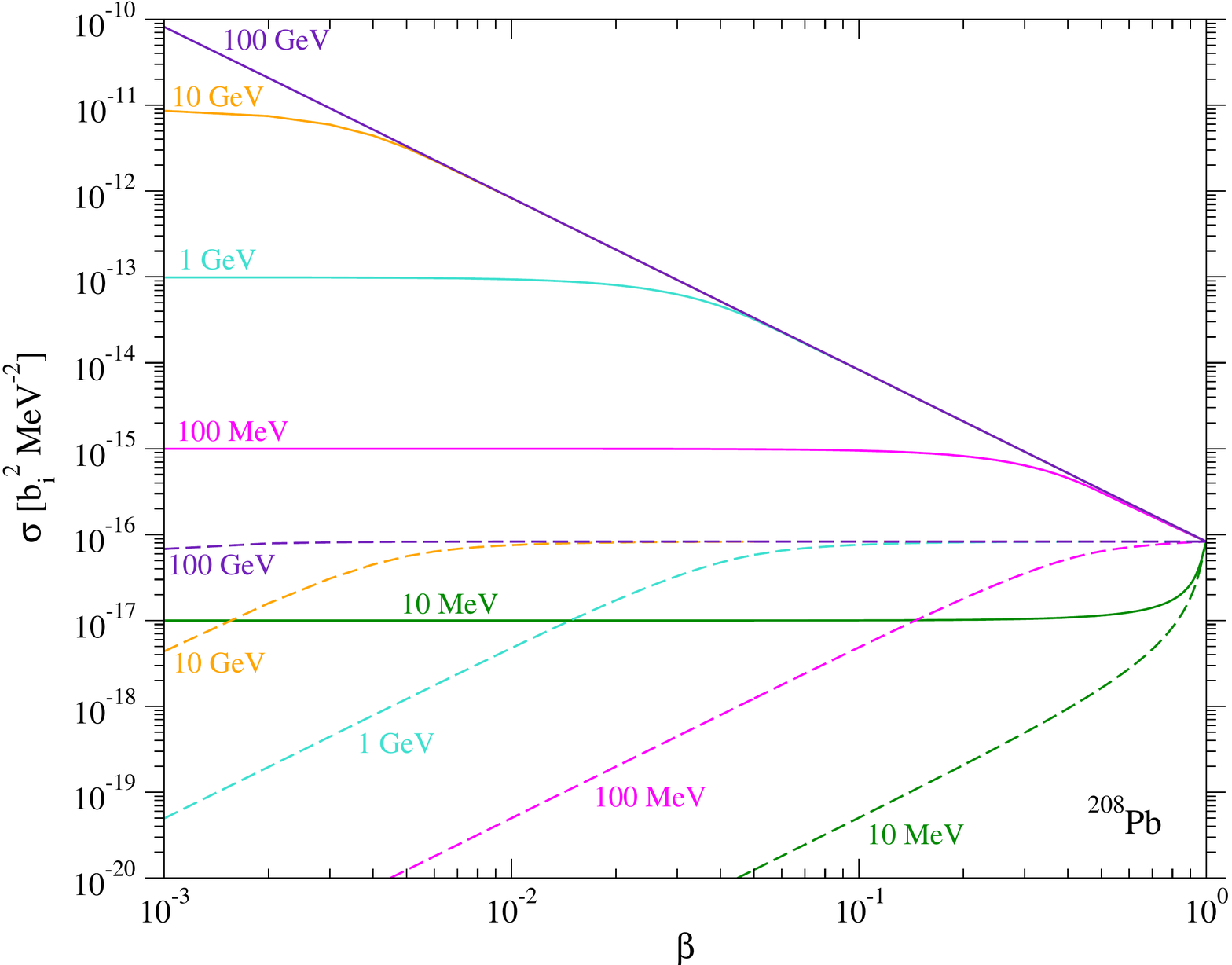}
\caption{(Color online) As for Fig. \ref{xs_vs_vel_c12}, but now for the integrated cross sections of the WIP-$^{208}$Pb interaction.
\label{xs_vs_vel_pb208}}
\end{center}
\end{figure}

In Fig. \ref{xs_vs_vel_c12} we show the vector contributions (solid lines) and the axial contributions (dashed lines) to the integrated WIP-$^{12}$C coherent cross section as a function of the WIP velocity for different WIP masses, from 1 eV to 10 MeV (to the left) and from 10 MeV to 1 TeV and above (to the right, with logarithmic scale in the abscissa). As before, they are given in units of the WIP couplings $b_V^2$ (vector contributions) or $b_A^2$ (axial contributions) and using the SM WNC values for the overall and hadronic couplings ($Z^0$ exchange). Similar results are shown in Fig. \ref{xs_vs_vel_pb208}, but now for a $^{208}$Pb target.

As can be seen in these figures, for WIP masses below 100 MeV the vector contributions (solid lines) remain within the same order of magnitude for $\beta$ below 0.9, and a squared dependence of the cross section on the WIP mass is clear in that region. The axial contributions show in the same region a squared dependence on the velocity parameter $\beta$ in addition to the mass squared dependence. These features, together with the squared target mass dependence due to coherent dominance of the scattering, allow us to write the vector and axial cross sections for $\beta\lesssim$ 0.5 and \mbox{$m \lesssim$ 100 MeV} simply as:
\begin{eqnarray}
\sigma_V\:[\text{MeV}^{-2}] &\approx& 2.64 \cdot 10^{-30} \:\left(M[\text{MeV}]\right)^2 \:\left(m[\text{MeV}]\right)^2 \:b_V^2 \\
\sigma_A\:[\text{MeV}^{-2}] &\approx& 1.32 \cdot 10^{-30} \:\beta^2 \:\left(M[\text{MeV}]\right)^2 \:\left(m[\text{MeV}]\right)^2 \:b_A^2 =	\frac{\beta^2}{2} \:\frac{b_A^2}{b_V^2} \:\sigma_V \;.
\label{xs_approx}
\end{eqnarray}

For the same WIP mass range but velocities closer to the speed of light there is a fast increase in the vector cross section, whereas for WIP masses larger than 100 MeV they decrease significantly as the WIP velocity increases. There is a smooth transition from small to large WIP mass behaviors, and the curves coalesce for masses larger than a few hundred GeV. As for the axial contributions (dashed lines), they increase with the WIP velocity. For masses below 100 MeV they show a fast increase in the WIP velocity ranges 0 $\lesssim\beta\lesssim$ 0.1 and 0.9 $\lesssim\beta\lesssim$ 1, with different sign of the second derivative in each region. In between, a smooth slope accounts for an increase of approximately two orders of magnitude. For masses above 100 MeV the increase in the 0.9 $\lesssim\beta\lesssim$ 1 region disappears, and the curves coalesce into an approximately constant curve above $\beta=$ 0.1. 

\begin{figure}
\begin{center}
\includegraphics[trim= 0cm 0cm 2cm 2cm, clip, width=0.48\textwidth]{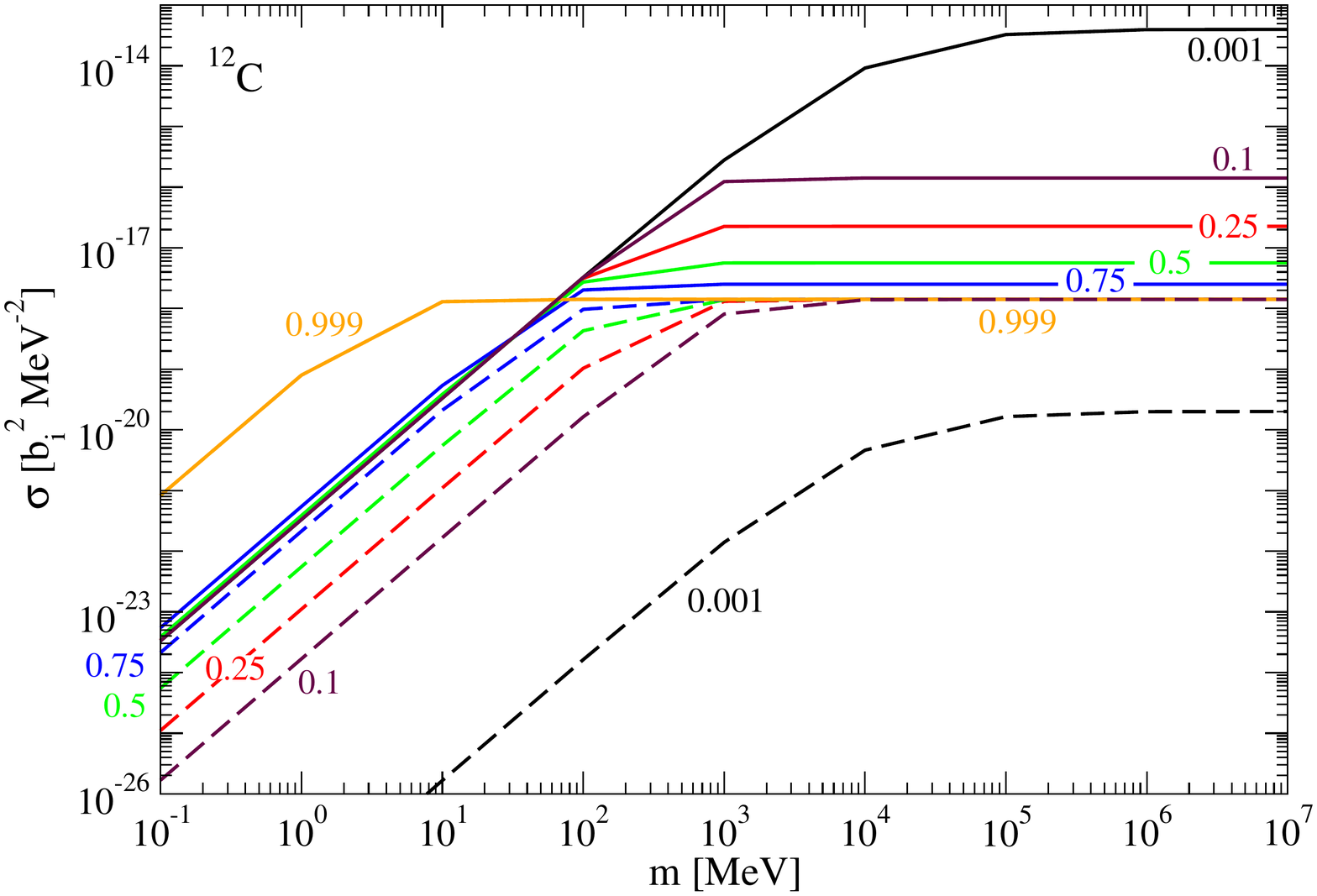}
\hspace{0.5cm}
\includegraphics[trim= 0cm 0cm 2cm 2cm, clip, width=0.48\textwidth]{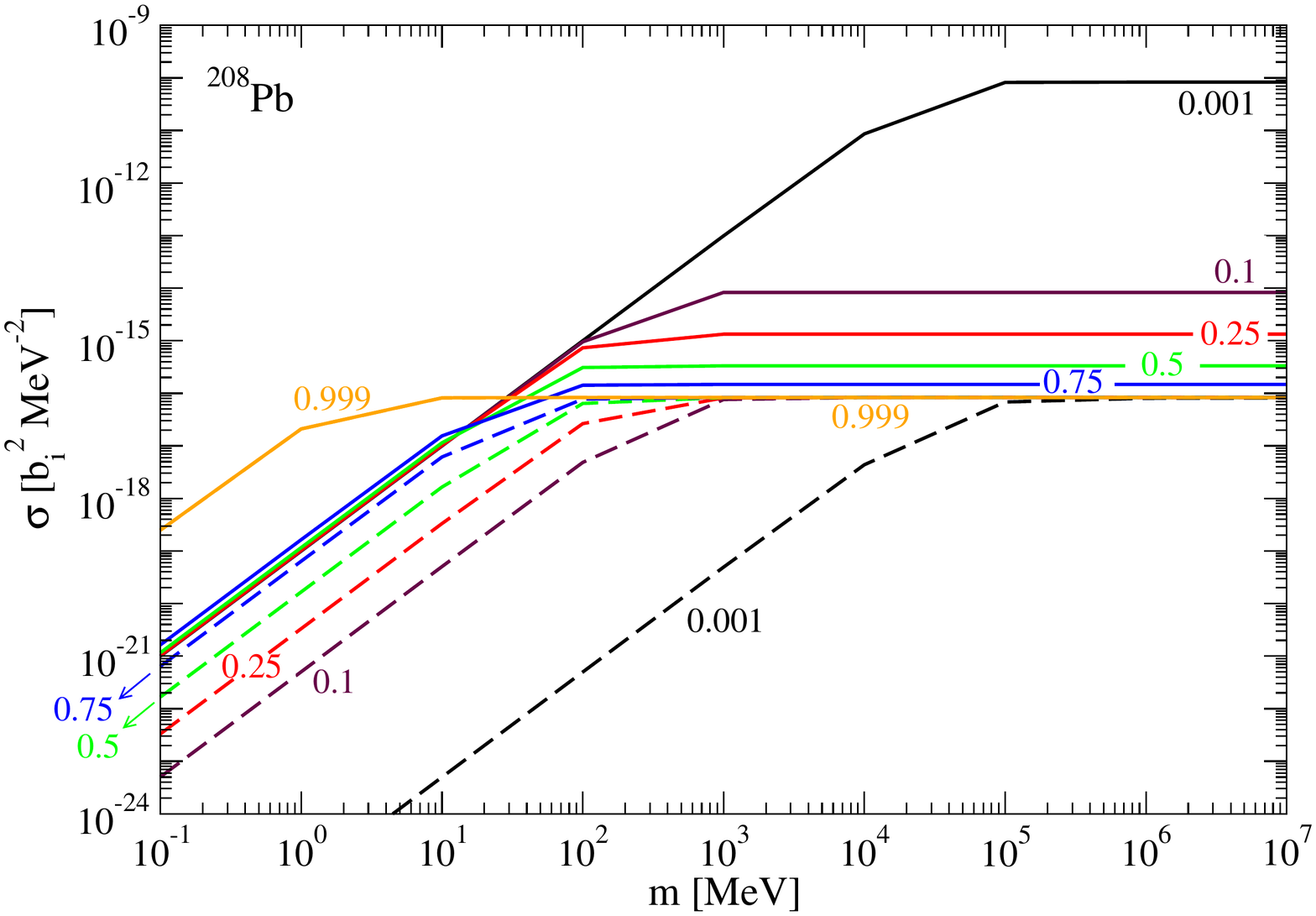}
\caption{(Color online) Vector (solid lines, in units $b^2_V$ MeV$^{-2}$) and axial (dashed lines, in units $b^2_A$ MeV$^{-2}$) contributions to the integrated cross sections of the WIP-nucleus interaction through $Z^0$ exchange, as a function of the WIP mass for different WIP velocities in terms of $\beta$ (given next to each curve). Left panel: WIP-$^{12}$C cross sections. Right panel: WIP-$^{208}$Pb cross sections.
\label{xs_vs_m}}
\end{center}
\end{figure}

The WIP mass dependence of the cross sections is more clearly seen in Fig. \ref{xs_vs_m}; several curves for fixed WIP velocity $\beta$ are shown for a $^{12}$C target (to the left) and for a $^{208}$Pb target (to the right). It is apparent how the curves corresponding to the vector contribution coalesce for $\beta\lesssim 0.9$ and $m\lesssim 100$ MeV, and the ones corresponding to axial contributions do so for $\beta\gtrsim 0.9$ and $m\gtrsim 1$ GeV.

\section{Conclusions}\label{conclusions}
We have studied the elastic scattering of weakly interacting particles (WIPs) off nuclei through vector and axial currents with massive boson(s) exchange, such that the momentum transfers are much lower than the boson(s) mass, $M_B >> Q$. We have isolated the dominant contribution to the elastic cross section, namely the coherent term, which is purely vector isoscalar on the nuclear vertex (and thus spin-independent). On the projectile vertex, both vector and axial contributions are considered separately. We have computed coherent cross sections for a wide range of WIP masses and velocities, as well as arbitrary WIP-boson vector and axial couplings ($b_V$ and $b_A$), arbitrary nucleon-boson vector isoscalar coupling ($\gamma_{V(0)}$), and arbitrary overall boson coupling ($\kappa$). The vector-current contribution to the WIP-nucleus cross section is therefore proportional to $b_V^2\:\gamma_{V(0)}^2\:\kappa^2$, whereas the axial-current contribution is proportional to $b_A^2\:\gamma_{V(0)}^2\:\kappa^2$. In the SM this process corresponds to active neutrino WNC scattering off nuclei (through $Z^0$ exchange), where both the vector- and the axial-current parts of the cross section are proportional to $a_{V,A}^2\:\beta_{V(0)}^2\:G_F^2=2.9\cdot10^{-23}$ MeV$^{-4}$.  

Results have been obtained for a light $N=Z$ (isospin-0, except for small isospin mixing) target, $^{12}$C, and a heavy $N>>Z$ target, $^{208}$Pb. First, nuclear Coulomb form factors have been shown for each target as a function of the momentum transfer; then the vector and the axial WIP longitudinal Rosenbluth factors have been given for different WIP velocities, again as a function of the momentum transfer. Using these calculations we have then obtained the vector and the axial contributions to the differential coherent cross sections for both nuclear targets as a function of the momentum transfer. And finally, we have shown the integrated cross sections for several WIP masses as a function of the WIP velocities. We have found that for WIP masses below 100 MeV the vector contributions are kept within the same order of magnitude for $\beta$ below 0.9, whereas the axial contributions show in the same region a squared dependence on the velocity parameter $\beta$; in both cases there is a WIP mass squared dependence, as well as a target mass squared dependence due to coherence. As the WIP velocity approaches the light limit the vector cross section increases rapidly when the WIP mass is below 100 MeV, but it decreases when the WIP mass is above approximately 100 MeV. The curves change smoothly as the WIP mass increases, particularly in the 100 MeV region. As for the axial contributions, for masses below 100 MeV they increase with the WIP velocity, rapidly in the regions 0 $\lesssim\beta\lesssim$ 0.1 and 0.9 $\lesssim\beta\lesssim$ 1, and smoothly in between; for masses above 100 MeV the increase in the 0.9 $\lesssim\beta\lesssim$ 1 region disappears, and the curves coalesce into an approximately constant curve above $\beta=$ 0.1. The  vector and axial integrated cross sections have also been plotted against the WIP masses for several WIP velocities, to show more clearly the WIP mass dependence.

Using the coherent WIP-nucleus cross sections obtained above it is possible to estimate the number of events of this kind expected in a given detector. The number of detections per year per ton of a given target material can be estimated as: 
\begin{eqnarray}
{\mathcal N} &\approx& 7.4 \times 10^{15} \:\frac{1}{A} \:F[\text{s}^{-1}\text{cm}^{-2}] \:\sigma[\text{MeV}^{-2}] \;,
\label{events}
\end{eqnarray}
where $F$ is the WIP flux, given in s$^{-1}$cm$^{-2}$, $\sigma$ is the WIP-target coherent cross section, given in MeV$^{-2}$ as in the results of the previous section, and $A$ is the mass number of the target material. For example, for the cross sections of Eqs. (\ref{xs_approx}), the number of expected detections per year per ton is:
\begin{eqnarray}
{\mathcal N}_V &\approx& 6.8 \times 10^{8} \:\frac{1}{A} \:\left(M[\text{MeV}]\right)^2 \:\left(m[\text{MeV}]\right)^2 \:b_V^2 \:\gamma_{V(0)}^2 \:\left(\kappa[\text{MeV}^{-2}]\right)^2 \:F[\text{s}^{-1}\text{cm}^{-2}] \\
{\mathcal N}_A &\approx& 3.4 \times 10^{8} \:\frac{1}{A} \:\beta^2 \:\left(M[\text{MeV}]\right)^2 \:\left(m[\text{MeV}]\right)^2 \:b_A^2 \:\gamma_{V(0)}^2 \:\left(\kappa[\text{MeV}^{-2}]\right)^2 \:F[\text{s}^{-1}\text{cm}^{-2}] \;.
\end{eqnarray}
The latter expressions are not given for a $Z^0$ exchange, but instead for a more general boson with vector isoscalar hadronic coupling $\gamma_{V(0)}$ and overall coupling $\kappa$, whose values are to be provided by a particular WIP theory. The overall coupling in the previous expressions must be introduced in MeV$^{-2}$ (since it is an effective coupling, for $M_B>>Q$) and the WIP and target masses in MeV.

Since the WIPs are supposed to be very elusive particles, their elastic interaction with nuclei can only be detected through the nuclear recoil caused by the scattering; the same happens with the SM neutrinos in elastic neutral current scattering from nuclei. Therefore, in addition to the expected number of events, it is also important to know the expected nuclear recoil, since very small values are very hard to detect. However, recoil energy thresholds as low as a few keV might be possible in the near future for a wide variety of detection techniques (scintillation, ionization, tracking, bubble chamber, {\it etc.}) and materials (noble liquids, semiconductor solids, {\it etc.}); masses of target materials might also reach in the near future the ton scale \cite{schol06}. From energy conservation, the recoil energy of the nuclear target is equal to the energy carried by the boson, namely the energy transfer. We have given results for differential cross sections as a function of the momentum transfer $q$, that can be easily translated into recoil energy using the condition of elastic scattering (Eq. (\ref{elastic_condition})): $E_{rec}=\omega=\sqrt{M^2+q^2}-M$.

\acknowledgments
We thank E. Moya de Guerra for useful comments on the manuscript. O.M. acknowledges support from a Marie Curie International Outgoing Fellowship within the European Union Seventh Framework Programme, under Grant Agreement PIOF-GA-2011-298364 (ELECTROWEAK), and MINECO FIS2011-23565 and FIS2014-51971-P. T.W.D. is supported in part by the Office of Nuclear Physics of the U.S. Department of Energy under Grant Contract No. DE-FG02-94ER40818.

\end{document}